\documentclass[12pt]{amsart}
\usepackage{amssymb,enumerate}
\usepackage{amsmath}
\usepackage{graphics}
\usepackage{cjmb5}
\usepackage{color}
\numberwithin{equation}{section}
\usepackage{wrapfig}
\usepackage{multirow}
\usepackage{listings}
\usepackage{algpseudocode}
\input{stefmac.sty}

\begin{document}
\title{On contour representation of two dimensional patterns}
\author{I.T. Banu-Demergian and G. Stefanescu}
\address{
Department of Computer Science\newline\indent 
University of Bucharest, Romania
}
\email{\newline\indent
iulia.banu@fmi.unibuc.ro\newline\indent
gheorghe.stefanescu@fmi.unibuc.ro
}
\setcounter{page}{1}
\coordinates{xx}{2011}{1}{00-00}
\date{date}
\undate{date}
\subjclass[2010]{code, code.}
\keywords{\em regular expressions, two-dimensional patterns, contours, structured interactive programming, formal methods}

\begin{unabstract}
Two-dimensional patterns are used in many research areas in computer science, ranging from image processing to specification and verification of complex software systems (via scenarios). The contribution of this paper is twofold. First, we present the basis of a new formal representation of two-dimensional patterns based on contours and their compositions. Then, we present efficient algorithms to verify correctness of the contour-representation. Finally, we briefly discuss possible applications, in particular using them as a basic instrument in developing software tools for handling two dimensional words.  
\end{unabstract}
\maketitle

\section{Introduction} 

The study of two-dimensional shapes is of wide interest. Applicability in pattern recognition, image processing, computer graphics and, more recently, in interactive computation demonstrates the need of a compact and steady model to handle two-dimensional objects. Contours-based representations are often used as they fit efficiency and simplicity requirements. Among them, chain-codes allow compression, without losing any information. The general idea is to encode the border of an image by a list of line-segments characterized by length and direction. The first chain-code representation is due to Freeman, 1961 \cite{free}. It describes a curve by linking adjacent points by one of eight possible moves, corresponding to an $i * 45^\circ$ angle, $i=0 .. 7$. Others encoding schemes, based on Freeman codification, have been proposed in \cite{liu05,koku85,bib01}. Kaneko and Okudaira \cite{koku85} obtained a high compression rate in coding geographic maps, using the property that a curve with gentle curvature is divided into long curve segments, each of which being represented by a sequence of two adjacent chain codes. E. Biribiesca \cite{bib01} presented a formal language approach, specifying some algebraic properties of chain codes representing 3D curves.

We propose a chain code based on four directions, over rectangular grids. Each line segment in the boundary of a two-dimensional image is identified with a letter in the set $\{u,d,r,l\}$ ($u$ stands for ``up'', $d$ for ``down'', $r$ for ``right'', and $l$ for ``left''), followed by a number, denoting its length. Roughly speaking, a {\em contour} is a closed line formed by a list of connecting segments, starting at a particular point, surrounding a finite internal area; a contour is associated to a bi-dimensional shape by a left-handed traversal. A general \emph{2-dimensional word} is specified by filling the area delimited by a contour with letters form a given alphabet.

For (1-dimensional) words, a powerful representation is provided by finite automata, regular expressions, and Kleene algebras. The connection between Kleene algebras and finite automata \cite{conway71} is well known and it provides a rich support for many others semantic models of computation, including models of parallel systems as: tile systems \cite{gi-re97}, Petri nets \cite{ga-ra92}, timed automata  \cite{acm02}, etc. A formalism for interactive parallel computation rv-IS (register-voice interactive systems) and a core programming language Agapia have been recently introduced in \cite{ste06a,dr-st08b}. They are based on finite interactive systems, a 2-dimensional version of finite automata. Using register machines and space-time duality, the formalism responds to the growing need of programming and reasoning about interactive systems. Its semantics is given in terms of scenarios, built up on top of 2-dimensional words.

The set of contours is enriched with a collection of composition operators. The obtained formalism allows defining a new type of regular expressions over 2-dimensional words n2RE \cite{bps13}, similar to 1-dimensional Kleene formalism. Many interesting open problems naturally occurs in this new formalism n2RE. Here, we are dealing with the formal representation of contours and efficient algorithms for contour representation correctness.

The contribution of this paper is twofold. First, we describe a formal representation of two-dimensional patterns based on contours and their compositions. Then, we present efficient algorithms to verify correctness of the contour-representation. 

\section{Arbitrary shapes in the 2-dimensional plane}

\subsection{Contours}

\begin{figure}[h]
\begin{tabular}{cc}
\raisebox{.3cm}{\includegraphics[scale=.35]{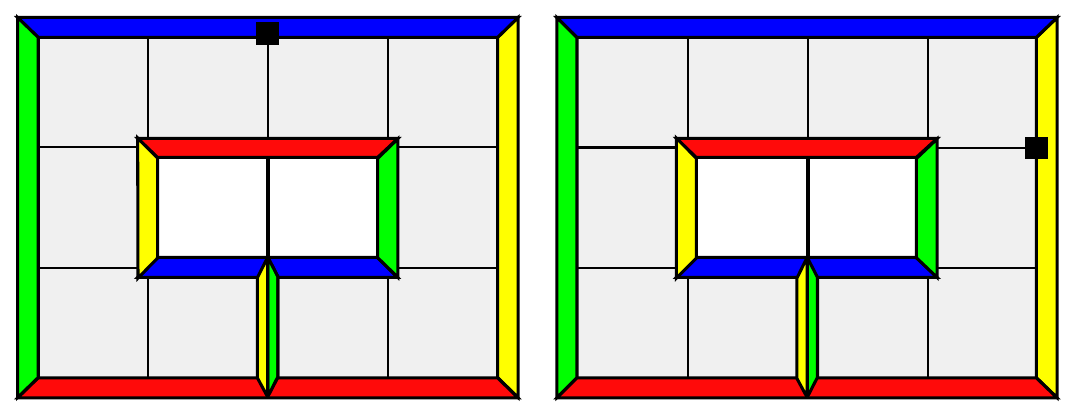}} 
& \includegraphics[scale=.35]{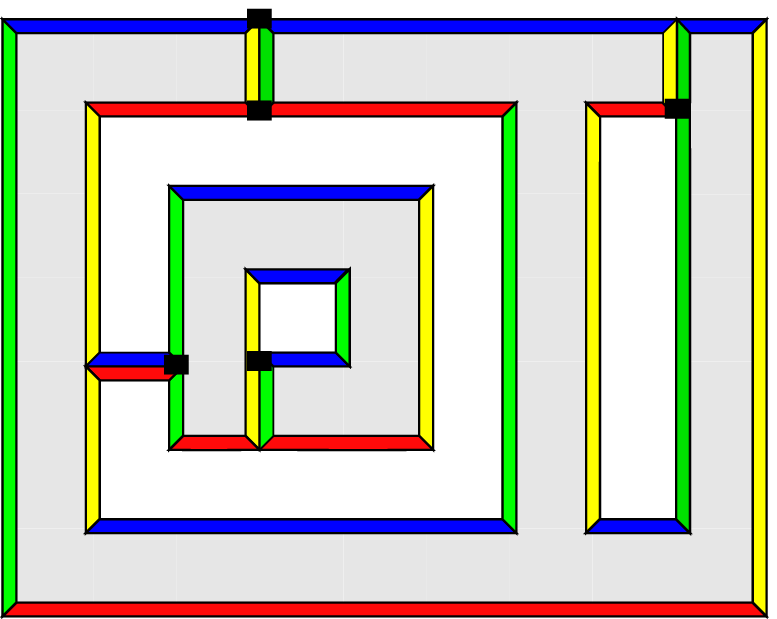}\\
(a)\hspace{4cm}(b) & (c)
\end{tabular}\vspace{-0.3cm}
\caption{Contours}
\label{expcont}
\end{figure}

A \textit{(pointed) contour} is a closed, non-overlapping line on a rectangular grid, $\mathbb{Z} \times \mathbb{Z} $, with a chosen start point, surrounding a finite internal area. Each of its segments will be represented using a letter from the set $\{u,d,r,l\}$ ($u$ stands for ``up'', $d$ for ``down'', $r$ for ``right'', and $l$ for ``left''), followed by a number denoting its length. A few examples of contours are shown in Fig.~\ref{expcont}.

A contour encloses disjoint interior \textit{components}, linked via empty shapes as $rrudll$; the (sub)contours surrounding empty shapes and travelling into the internal area are named \emph{tunnels}, while those in external areas are called \emph{bridges}. A clockwise traversal determines the 2-dimensional area associated to a contour. The area on the east side of a $u$ move is internal, while the one on the west is external. Similar conventions hold for $r,d$ and $l$. Multiple surrounding of the same zone as well as infinite internal areas are forbidden. 

Two contours are equivalent iff they enclose the same internal area, modulo translations. For instance, a different placement of the start point determines an equivalent circularly shifted representation. Two equivalent contours are $rrdddllurulldrdlluuurr$ (shortly written as $r^2d^3l^2urul^2drdl^2u^3r^2$) and $d^2l^2urul^2drdl^2u^3r^4d$ - they are presented in Fig.~\ref{expcont}(a),(b). 
By filling the interior area of a contour with letters from a given alphabet one gets a \textit{general 2-dimensional word}.

In preparation for the forthcoming formal definition of a contour, some more notations are needed.

\textit{Line segments:} For a vertical line segment $l=((x,y),(x,y+1))$ (simply denoted as $l = (x,y+0.5)$), we denote by $l^k_C$ a predicate which is true if and only if the difference between the "up" and "down" arrows of $C$ passing over $l$ is $k$; notice that $k \in \mathbb{Z}$. A similar notation is used for horizontal line segments $l=((x,y),(x+1,y))$. 

\textit{Cells:} For a cell $\{(x,y),(x+1,y),(x+1,y+1),(x,y+1)\}$, represented by its center point $c=(x+0.5,y+0.5)$, we denote by $c^k_{C,w}$ a predicate which is true if and only if $l^k_C$ is true, where $l$ is the first line  covered by $C$ and situated on the west side of $c$; in this counting bridges and tunnels (lines having equal up/down passings) are skipped. Formally, let\\
\hspace*{.5cm}
$z=max\{w\in \mathbb{Z}: w\leq x$ and ($l=(w,y+0.5)$ is such that $l^k_C$ is true for a $k\neq 0)\}$;\\
then $c^k_{C,w} = lz^k_C$, where $lz = (z,y+0.5)$.

\textit{Internal points:} A cell $c$ is \emph{internal from a west perspective} if $c^k_{C,w}$ is true for a $k > 0$, meaning there are more $u$ than $d$ passings of the first line segment found towards the west, ignoring bridges and tunnels. The notations $c^k_{C,e}$, $c^k_{C,n}$, and $c^k_{C,s}$ are similarly used to define internal cells from the other perspectives, i.e., examining respectively the est, the north, and the south neighbourhood. The set of cells which are internal to a contour $C$ from all directions is denoted by $Int(C)$.

\textit{External points:} A cell is external from a west perspective if, going towards west, there is no segment with unbalanced $u/d$ moves travelled by $C$ or  there is a first line segment with unbalanced $u/d$ passings via $C$ and having more $d$ than $u$ moves. Formally, for all $k$, either $c^k_{C,w}$ is false or $ c^k_{C,w}$ is true for a $k<0$. $Out(C)$ denotes the set of cells which are external from all directions.

A contour is well-defined if the set of interior cells $Int(C)$ is finite and the intersection $Int(C) \cap Out(C)$ is empty. The precise definition is described below.

\begin{definition}\label{interiorpoints}
(a) A string over $\{u,d,l,r\}$ represents a \emph{valid contour} if it describes a closed line and any cell is either internal from all directions or external from all directions; moreover, for all internal cells $C$, all $ c^k_{C,w},  c^k_{C,n}$ are satisfied with $k = 1$ and  all $c^k_{C,e}, c^k_{C,s}$ are satisfied with $k = -1$.

(b) Two contours $C1$ and $C2$ are considered \emph{equivalent} if and only if $Int(C1) = Int(C2)$.
\end{definition}

In order to avoid overlapping, each internal cell is surrounded only once. For instance $rdlurdlu$ in not a valid contour, while $rdlu$ is. This shows why in the above definition $|k|$ is restricted to be $1$.\svsp

When deciding if a string represents a valid contour, it is useful to have a set of criteria dealing with contour segments, not with the cells. Indeed, inspecting all cells in the grid can be algorithmically inefficient. The equivalent definition in Prop.~\ref{equivalentdef} below will be used by the next section algorithms to check if a contour is well defined. Finiteness of $Int(C)$ is equivalent to conditions \ref{closedline} and \ref{closedshape}. Conditions \ref{repeatparse} and \ref{intersections} ensure that $Int(C) \cap Out(C)$ is empty.

\begin{proposition}\label{equivalentdef}
Let $lv$ (resp. $lh$) denote vertical (resp. horizontal) line segments of a string $C\in\{u,d,r,l\}^*$ enriched with a start point. Then, $C$ represents a valid contour if and only if the following conditions are satisfied:

\bd\item[(closed line)]
\begin{equation}\label{closedline}
\sum_{lh_C^k}k = 0 \hsp and\hsp \sum_{lv_C^k}k = 0
\end{equation}\snvsp

\item[(closed shape)]\snvsp
\begin{equation}\label{closedshape}
\begin{split}
& \forall x: \text { let } y_x=min\{y\in \mathbb{Z}: \exists  l  =(x + 0.5,y) \text { such as } l_C^k \text{ is satisfied for a } k \neq 0 \};\\ 
& \text{if } y_x \neq nill \text{ then the corresponding } l_C^k \text{ is true for a } k < 0.
\end{split}
\end{equation}\snvsp
 
\item[(no repetitions)]\snvsp
\begin{equation} \label{repeatparse} 
\forall l: \text{ if } (l=(x,y + 0.5) \text{ or } l = (x+0.5,y)) \text{ and } l_C^k \text{ is true for a } k \neq 0, \text{ then } k \in \{1,-1\}.
\end{equation} \snvsp

\item[(alternation in-out)]\snvsp
\begin{equation}\label{intersections}
\begin{split}
& \text{ for any pair }l_1 = (x + 0.5, y_1), l_2 = (x + 0.5, y_2) \text{ of consecutive horizontal borders }\\
& (\text{that is,}\ \forall l = (x + 0.5, y): \text{ if } y_1 < y <y_2, \text{ then } (l_C^k \text{ true }\Rightarrow (k = 0)))\\
& \text{ we have: } if (l_1)_C^{k_1} \text{ and } (l_2)_C^{k_2} \text{ are true, then } k_1 + k_2 = 0.
\end{split}
\end{equation}\snvsp
\ed\snvsp
\end{proposition}

\textit{Comments:} 
Condition 2.1 says the number of left moves equals the number of right moves; and similarly for the vertical direction.

By 2.2, the horizontal line segment with the lowest $y$ coordinate (a line segment situated to the extreme south border) must be oriented left to right. This condition ensures the internal area be finite. For instance $drul$ is not a valid representation, violating this condition.  Equivalent presentations of this condition may be introduced using the other directions.

Condition 2.3 has easy intuitive meaning: a contour has no repeated parsing on the borders of a non-empty internal area.

Finally, 2.4 says a contour has no self-intersection, except for tangential contact of disjoint areas, namely tunnels. Horizontal segments with unequal $r/l$ passing, situated at the same $x$ coordinate, must alternate $r/l$ directions (i.e., the difference $r-l$ is a sequence $-1,1,-1,\dots$ or $1,-1,1,\dots$). This ensures each cell belongs either to $Int(C)$ or to $Out(C)$.\svsp

\begin{figure}[h]
\begin{tabular}{cccc}
\includegraphics[scale=0.30]{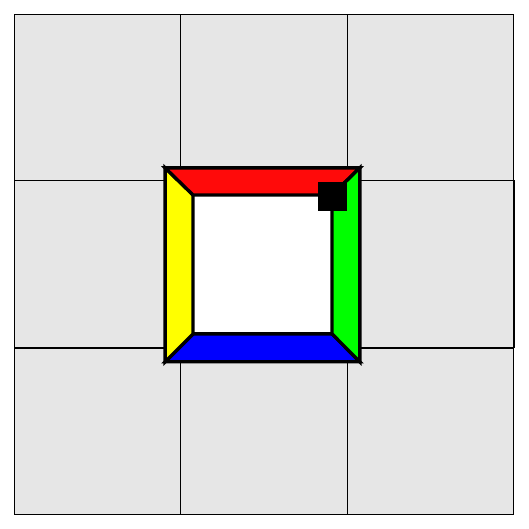} & \includegraphics[scale=.15]{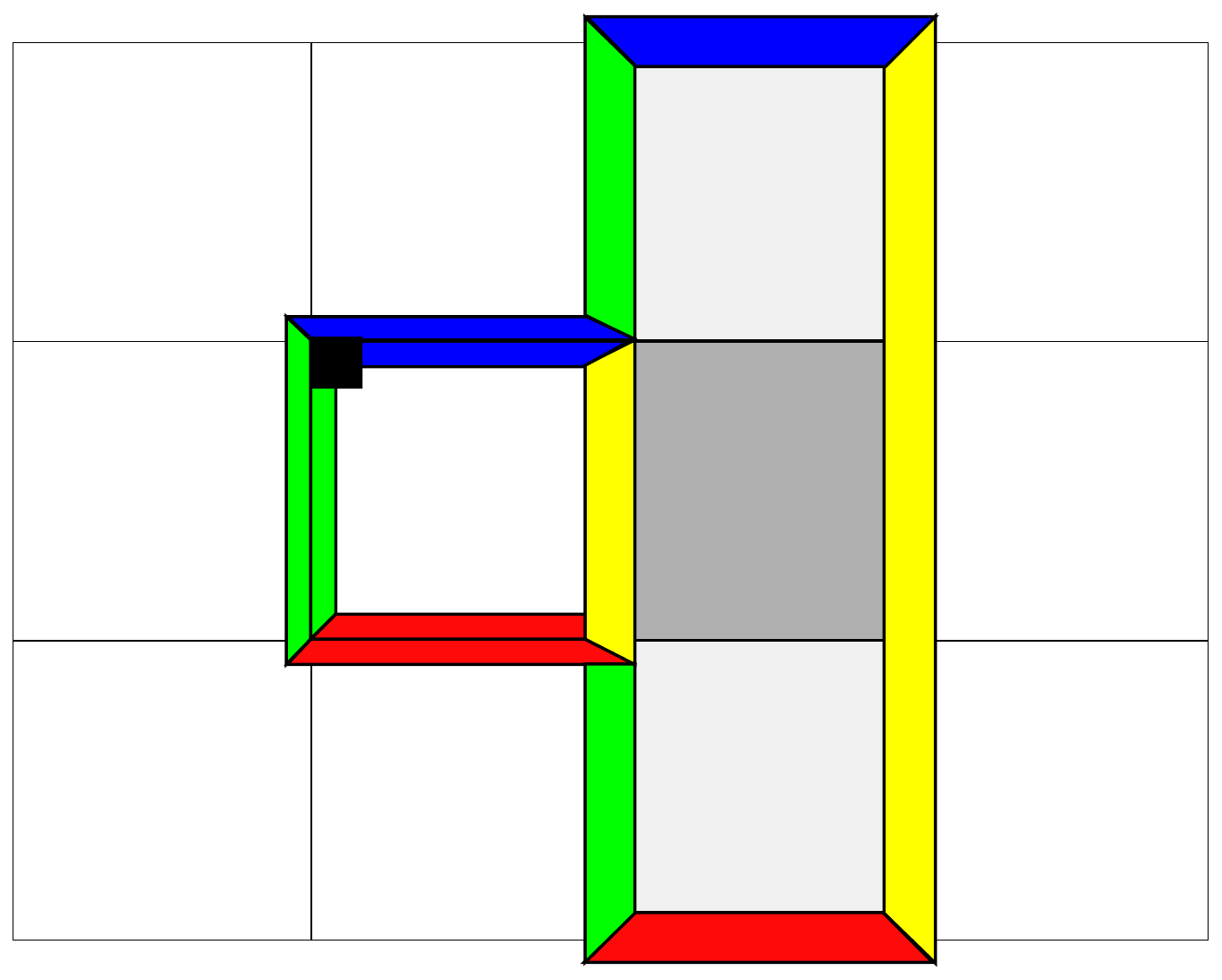} & \includegraphics[scale=.15]{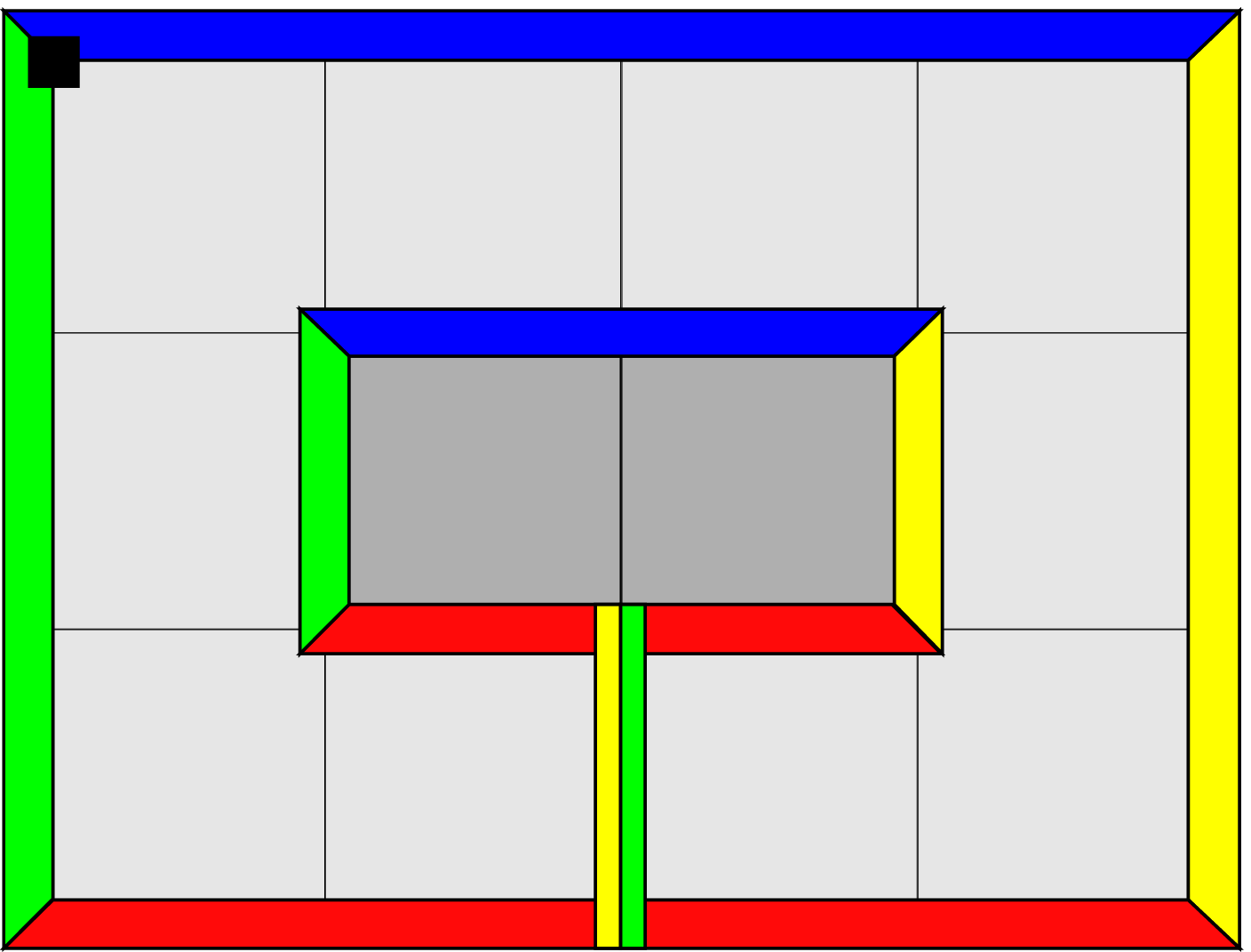}
& \includegraphics[scale=.15]{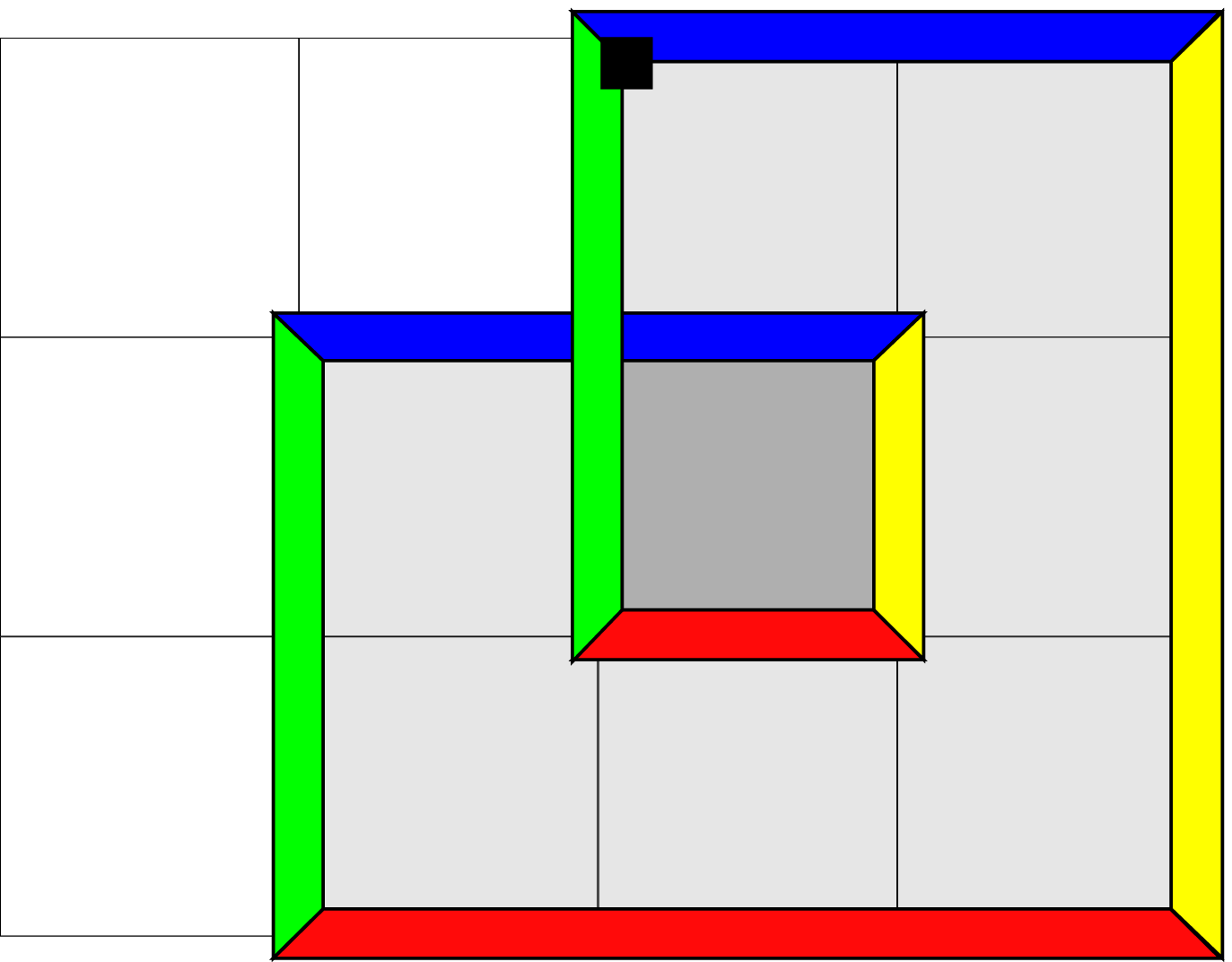}\\
(a) & (b) & (c) & (d) 
\end{tabular}\vspace{-.3cm}
\caption{Examples of invalid contours}
\label{fig:invcont}
\end{figure}

In Fig.~\ref{fig:invcont} some examples of invalid contour representations are illustrated: (a) $ldru$, (b) $rdlururd^3lulu$, (c) $r^4d^3l^2ulur^2dldl^2u^3$, and (d) $r^2d^3l^3u^2l^2dlu^2$. The contour in (a) is not representing a finite shape; the contour in (b) is passing the cell $(1.5, 1.5)$ twice; the cells $(1.5,1.5)$ and $(2.5, 1.5)$ belong either to the interior or to the exterior area of the contour in (c); finally, the contour in (d) is self-intersecting.

\out{
It can be seen that even more simplified version of  restriction conditions may be formulated in terms of larger segments. 
}

\subsection{A normal form representation of contours}
A string $C$ is called a {\em simple contour} if it  represents a contour and there are no proper substrings of $C$ with this property. It represents a finite area with no holes. The representation of a simple contour is unique modulo the position of the start point.

Two \emph{edge-neighbouring cells} are two cells which share a horizontal or a vertical edge. An \emph{edge-connected component} is a maximal set of cells such that any two cells are connected with a path of edge-neighbouring cells, all from that component.

A (general) contour may be decomposed such that each of its {\em edge-connected} components is represented by a simple contour (used for its external border) and zero, one or more ``inverses'' of simple contours\foo{The \emph{inverse} of a string C over the alphabet $\{u,d,l,r\}$ is obtained by replacing $u/d/l/r$ with $d/u/r/l$, respectively.} for its possible internal holes. This decomposition, called the {\em normal form representation} of a contour, is unique up to connecting identities. (Recall that identities are contours with empty interior area, e.g., tunnels or bridges.)

For instance the contour: $r^5dld^5ru^6rd^7l^9u^7r^3dl^2d^3ru^2r^3d^3l^2uruld^2luld^2r^5u^5l^3u$, depicted in Fig.~\ref{expcont}(c), may be decomposed into two connected components and has the following normal form representation: \bi
\item the components: (a) $\{$contour $\{r^9d^7l^9u^7\}$ and holes $\{l^6d^6r^6u^6\}, \{ld^5ru^5\}\}$; and (b) $\{$contour $\{r^3d^3l^3u^3\}$ and holes $\{ldru\}\}$; 
\item the information on the relative position of the internal holes in the surrounding contours given by connecting identities (not shown here).\ei 

\out{
\begin{figure}[h]
\centering
\includegraphics[scale=.12]{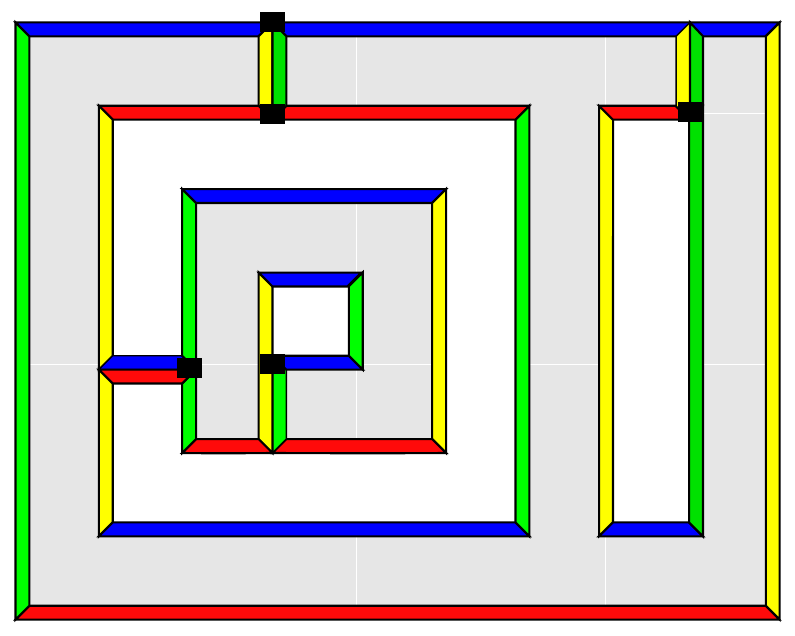}
\vspace{-.3cm} 
\caption{Contour with decomposition $\{\{r^9d^7l^9u^7\}$, $\{l^6d^6r^6u^6\}$, $\{ld^5ru^5\}$\}, $\{\{r^3d^3l^3u^3\}$, $\{ldru\}\}$}
\label{fig:normalfc}
\end{figure}
}

As illustrated by the example above, the normal form of a contour alternates the traversals of exterior and interior shapes. The only variation are the bridges/tunnels connecting simple contours. 

\subsection{2-dimensional regular expressions}

The set of well-defined contours is enriched with a binary composition operator ".": the result of composing $C1$ and $C2$ is the string $C1\ C2$, provided this is a valid contour. This means, the contours are gluing together via the starting points used in their representations.

For a graphical example, notice that C1 . C2 below shows a valid composition, while C2 . C3 shows an example of composition leading to an invalid result (the result has overlapping areas).
\bi\item[]\begin{tabular}{cc@{\hsp}ccc} C1 & C2 & C3	& C1 . C3 (valid) & C2 . C3 (not valid)\\ 
\includegraphics[scale=.35]{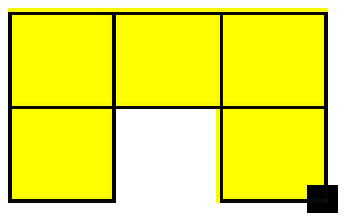} & \includegraphics[scale=.35]{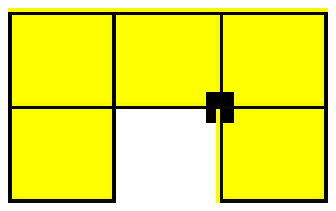} & \includegraphics[scale=.35]{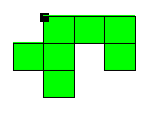} & \includegraphics[scale=.35]{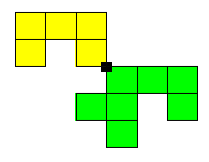} & \includegraphics[scale=.35]{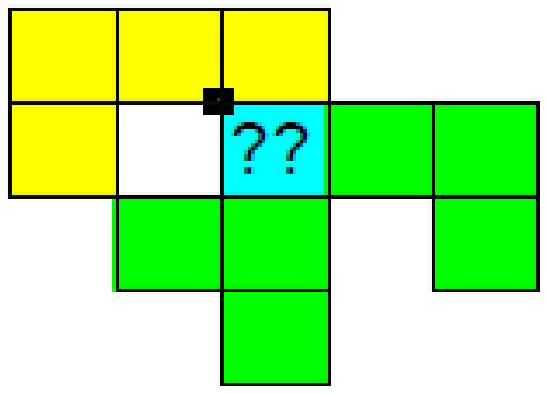}
\end{tabular}\ei 

\textit{Generic definition for restricted composition operators:} Restricted composition operators are obtained from the following generic format. Suppose we are given: 
\be\item 2 words $v,w$; a subset $Y$ of elements of the contour of $v$ (the yellow elements in the figure below); a subset $G$ of elements of the contour of $w$ (the green ones); the subset $B$ of actual contact elements after composing, as above, $v$ with $w$ via the points indicated by a little arrow (the blue elements).\\
\hspace*{2cm}\includegraphics[scale=.35]{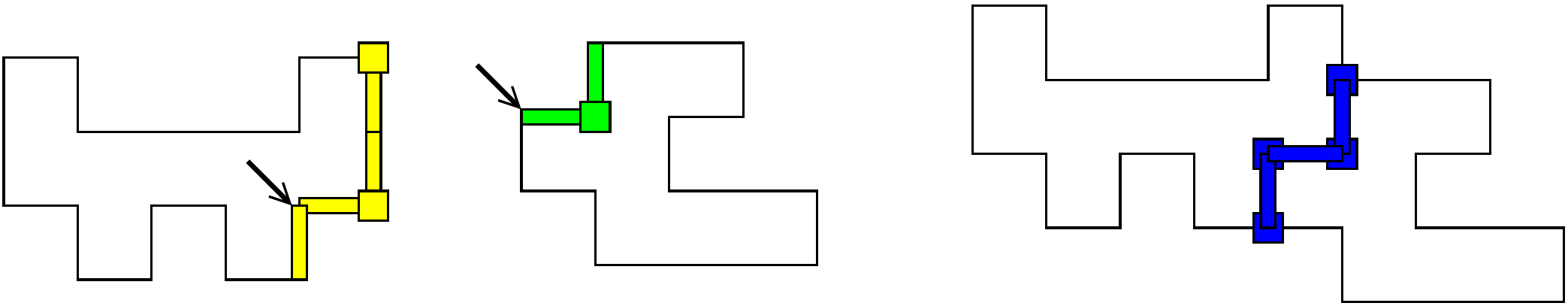}\\
\hspace*{2.8cm}$v$\hspace{2.2cm}$w$\hspace{1.5cm}$v\ R(Y,G,B)\ w$\item a relation $R(Y,G,B)$ between the above 3 subsets.
\ee
The resulted restricted composition is denotes by $v\ R(Y,G,B)\ w$.

In the given example, a relation $R$ making the restricted composition valid may be: $G\subseteq Y \wedge G\subseteq B$ (after composition, all the elements in the green set are on the common border and included in the yellow set). 

\textit{Example - A set of particular restricted composition operators 2RegExp:} A line $l=(x,y+0.5)$ is on the \textit{east border} of a contour $C$ if $l_C^k$ holds for $k = -1$; equivalently, the cell $(x-0.5,y+0.5)$ is internal, while  $(x+0.5,y+0.5)$ is in the exterior area. Similarly, a point $p=(x,y)$ is on the south-east border of $C$ if the cell $c=(x-0.5,y+0.5)$ is in the internal area of $C$, while the other 3 cells around are in the external area of $C$. Bridges or tunnels are not be counted as borders.

Let us use the following notation: $w$ for ``west border'', $e$ for ``east border'', $n$ for ``north border'', $s$ for ``south border'', $nw$ for ``north-west point'', $ne$ for ``north-east point'', $sw$ for ``south-west point'', and $se$ for ``south-east point''. We denote by $Connect$ their set $\{w,e,n,s,nw,ne,sw,se\}$. 

On each of the above eligible glueing combination $(x,y) \in Connect$ we put a constrain consisting of a propositional logic formula $F\in PL(\phi_1,\phi_2,\phi_3,\phi_4)$
\foo{$PL(Atom)$ denotes the set of propositional logic formulas built up with atomic formulas in $Atom$. For typing reasons, the boolean operations ``not'', ``and'',  and ``or'' are denoted by ``!'', ``\&'', and ``V'', respectively.}, i.e., a boolean formula built up starting with the following atomic formulas: 
\bi\item[] \hspace*{-.6cm}$\phi_1(x,y)={}$ ``$x < y$'', $\phi_2(x,y)={}$``$x = y$'', $\phi_3(x,y)={}$``$x > y$'', and $\phi_4(x,y)={}$``$x$~\#~$y$''.\ei
The meaning of the connectors is the following:  ``$<$'' - left is included into the right; ``$=$'' - left is equal to the right; ``$>$'' - left includes the right; ``$x$~\#~$y$'' - left and right overlaps, but no one is included in the other.

For instance: $f (e=w) g$ means ``restrict the general composition of $f$ and $g$ such that the east border of $f$ is identified  to the west border of $g$''; $f (e > w) g$ - the east border of $f$ includes all the west border of $g$, but some east borders of $f$ may still be not covered by west borders of $g$; etc. 

We also use the notation \bi\item[]$\phi_0(x,y)={}$``$x \ O \ y$'', where ``$O$'' means empty intersection.\ei Actually, this is a derived formula $\neg(\phi_1(x,y)\vee\phi_2(x,y)\vee\phi_3(x,y)\vee\phi_4(x,y))$.

\bdfn (restricted compositions) 
A \emph{restriction formula} $\phi$ is a boolean combination in $PL(F_1,\dots,F_n)$, where $F_i$ are constricting formulas  involving certain eligible glueing combinations $(x_i,y_i)\in Connect$. 
A \emph{restricted composition operation $\_(F)\_$} is the restriction of the general composition to composite words satisfying $F$. 
A word $h\in f~.~g$ belongs to $f\ (F)\ g$ if for all gluing combinations $(x_i,y_i)$ occurring in $F$ the contact of the $x_i$ border of $f$ and $y_i$ border of $g$ satisfies $F_i$.
\emph{Iterated composition operators} are denoted by $*(F)$, for a restriction formula $F$.\edfn

\textit{Example:} An example of 2RegExp expression (for spiral words \tdw{x} \tdw{2aa\\2x1\\bb1} \tdw{2aaaa\\22aa1\\22x11\\bbbb1} \dots) is:\vspace{-.4cm}\\
\verb+x(e<w & w<e & n<s & s<n)+\\
\verb+{[R(se>ne)D](nw<ne & sw<se)[L(nw>sw)U]}*_(e<w & w<e & n<s & s<n)+\\
where \verb+R = a*_(e.w),  D = 1*_(s.n),  L = b*_(e.w),  U = 2*_(s.n)+.

\section{Algorithms for testing correct representations of shapes}

In this section we presents two procedures for verifying the correctness of contour representations.\svsp

Dealing with cell criteria stated in Def. \ref{interiorpoints} may lead to inefficient algorithms. Basically, in order to decide if a given sting  $C\in\{u,d,r,l\}^*$ is a valid contour, one has to determine the membership of each cell either to $Int(C)$ or $Out(C)$ by calculating the predicates $c^k_{C,w},  c^k_{C,n}, c^k_{C,e}, c^k_{C,s}$. Thus, the advantage of having a 1-dimensional string representation is not exploit, as the analysis deals with the full 2-dimensional plane.

The following algorithms are based on the equivalent conditions in Prop.~\ref{equivalentdef}, dealing with contour segments. Testing if a contour is a closed line can be easily done in linear time; the main difficulty remains to verify conditions \ref{closedshape}, \ref{repeatparse}, and \ref{intersections}.

\subsection{The 1st algorithm for valid contours}

The first algorithm is based on sorting, achieving $O(n \log{n})$ complexity, where $n$ is the contour length.  The list $l[1 \dots n]$ of lines associated with the letters of the contour is sorted according to $(x,y)$ coordinates. The values $k$ that satisfy the predicates $l[i]_C^k$ are calculated in one traversal of the sorted list, as the informations to be added for each segment are situated on consecutive positions. Further checking of requirements \ref{closedshape}, \ref{repeatparse}, and \ref{intersections} is immediate. The full description of this 1st algorithm is shown in Fig.~\ref{alg:sort} and an example in Example~\ref{exp:asort}. It takes as input the string denoting the moves along the contour to be verified; the stating point is set to $(0,0)$ (the contours are invariant to translations, so the starting point position does not matter). 

\begin{figure}[h]
\centering
\begin{small}\begin{tt}
\begin{algorithmic}[1]
  \Function{TestContour}{char C[1 \dots n]}
\State int x, y, m; float aux[1 \dots n][1 \dots 3], int k, prec;  boolean valid;
\State x := 0, y := 0; 
\State \textbf{for }{k:=1 \textbf{to} n}
\State \hspace{0.15cm} \textbf{switch} C[k]
\State \hspace{0.3cm} \textbf{case} 'r': aux[k][1] := x + 0.5; aux[k][2] := y; x := x+1;
\State \hspace{0.3cm} \textbf{case} 'd': aux[k][1] := x; aux[k][2] := y - 0.5; y := y-1;
\State \hspace{0.3cm} \textbf{case} 'l': aux[k][1] := x - 0.5; aux[k][2] := y; x := x-1;
\State \hspace{0.3cm} \textbf{case} 'u': aux[k][1] := x; aux[k][2] := y + 0.5; y := y+1;
\State Sort(aux);
\State k:= 1; m:=0; valid:= true; prec := 0;
\State \textbf{while }{k $\le$ n}
\State \hspace{0.15cm}x := aux[k][1]; y := aux[k][2]; m:= m+1;
\State \hspace{0.15cm}l[m][1]:=x; l[m][2]:=y; l[m][3]:=0;
\State \hspace{0.15cm}\textbf{while }{k $\le$ n $\wedge$ x = aux[k][1] $\wedge$ y = aux[k][2]}
\State \hspace{0.3cm}\textbf{switch} C[aux[k][3]]
\State \hspace{0.45cm}\textbf{case} 'r': l[m][3] := l[m][3] + 1;
\State \hspace{0.45cm}\textbf{case} 'd': l[m][3] := l[m][3] - 1;
\State \hspace{0.45cm}\textbf{case} 'l': l[m][3] := l[m][3] - 1;
\State \hspace{0.45cm}\textbf{case} 'u': l[m][3] := l[m][3] + 1;
\State \hspace{0.3cm}k := k+1 ;
\State \hspace{0.15cm}\textbf{If }l[m][3] $<$ -1 $\vee$  l[m][3] $>$ 1
\State \hspace{0.3cm}valid := false;
\State \hspace{0.15cm}\textbf{If  }l[m][3] $<>$ 0 $\wedge$ prec $<>$ 0 $\wedge$ x = l[prec][1]
\State \hspace{0.3cm}\textbf{If  }l[prec][3] + l[m][3] $<>$ 0
\State \hspace{0.45cm}valid := false;
\State \hspace{0.3cm}prec:=m;
\State \hspace{0.15cm}\textbf{If  }(prec = 0 $\wedge$ round(x) $<>$ x) $\vee$
\\\hspace{1.7cm}(prec $<>$ 0 $\wedge$ x $<>$ l[prec][1] $\wedge$ round(x) $<>$ x)
\State \hspace{0.3cm}\textbf{If  }l[m][3] $>$ 0
\State \hspace{0.45cm}valid := false;
\State \hspace{0.3cm}\textbf{If  }l[m][3] $<>$ 0
\State \hspace{0.45cm}prec := m;
\State \textbf{return} valid;
  \EndFunction
\end{algorithmic}
\end{tt}\end{small}\vspace{-.3cm}
\caption{Algorithm for checking the correctness of contour representations}
\label{alg:sort}
\end{figure}

\bex \label{exp:asort}Taking as input the contour $C=rrrrdluurdlldluuld$ 
\begin{center}
\includegraphics[scale=.15]{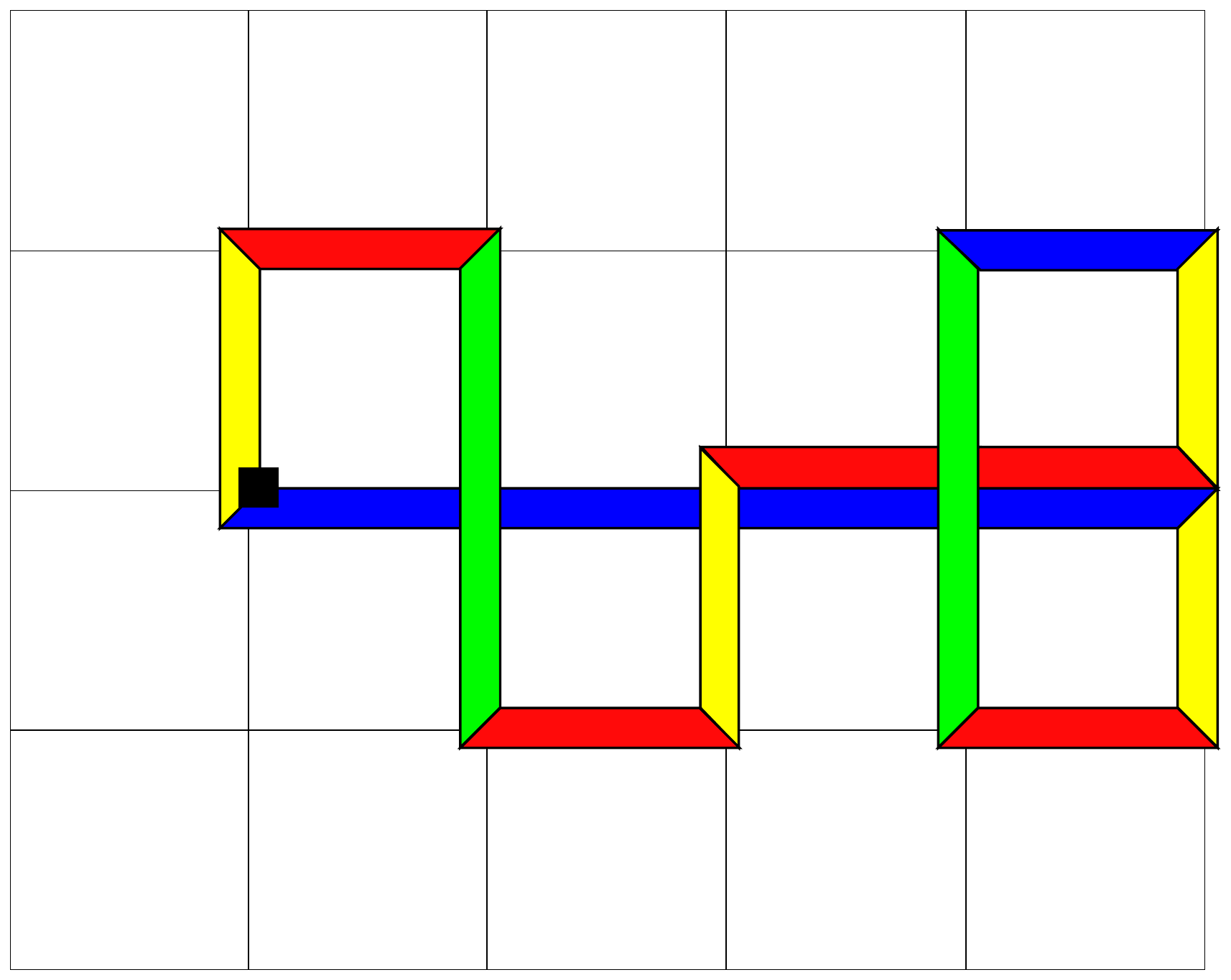}
\end{center}
the algorithm described in Fig.~\ref{alg:sort} runs as follows.\svsp

1. First it calculates the segments reached by $C$ and stores the result in \verb+aux+ (lines 4-9 in Algoritm \ref{alg:sort}).\svsp\\
\parbox[c]{13.7cm}{
$(0,0) \xrightarrow[(0.5,0)]{1:r}(1,0)
     	\xrightarrow[(1.5,0)]{2:r}(2,0) 
     	\xrightarrow[(2.5,0)]{3:r}(3,0) 
     	\xrightarrow[(3.5,0)]{4:r}(4,0)
	\xrightarrow[(4,-0.5)]{5:d}(4,-1) 
	\xrightarrow[(3.5,-1)]{6:l}(3,-1) 
	\xrightarrow[(3,-0.5)]{7:u}(3,0) 
	\xrightarrow[(3,0.5)]{8:u}(3,1) 
	\xrightarrow[(3.5,1)]{9:r}(4,1)
	\xrightarrow[(4,0.5)]{10:d}(4,0) 
	\xrightarrow[(3.5,0)]{11:l}(3,0) 
	\xrightarrow[(2.5,0)]{12:l}(2,0) 
	\xrightarrow[(2,-0.5)]{13:d}(2,-1) 
	\xrightarrow[(1.5,-1)]{14:l}(1,-1) 
	\xrightarrow[(1,-0.5)]{15:u}(1,0) 
	\xrightarrow[(1,0.5)]{16:u}(1,1) 
	\xrightarrow[(0.5,1)]{17:l}(0,1) 
	\xrightarrow[(0,0.5)]{18:d}(0,0)$
}\svsp

2. The second step is to sort the list above, according to $(x,y)$ coordinates:\svsp\\
\parbox[c]{13.7cm}{
$(0,0.5,\stackrel{18}{d})$ $(0.5,0,\stackrel{1}{r})$ $(0.5,1,\stackrel{17}{l})$ $(1,-0.5,\stackrel{15}{u})$ $(1,0.5,\stackrel{16}{u})$
$(1.5,-1,\stackrel{14}{l})$ $(1.5,0,\stackrel{2}{r})$ $(2,0.5,\stackrel{13}{d})$ $(2.5,0,\stackrel{3}{r})$ 
$(2.5,0,\stackrel{12}{l})$ $(3,-0.5,\stackrel{7}{u})$ $(3,0.5,\stackrel{8}{u})$ $(3.5,-1,\stackrel{6}{l})$ $(3.5,0,\stackrel{4}{r})$ $(3.5,0,\stackrel{11}{l})$ $(3.5,1,\stackrel{9}{r})$ $(4,-0.5,\stackrel{5}{d})$ $(4,0.5,\stackrel{10}{d})$
}\svsp

3. The final loop (lines 17-20) calculates how many times each segment is passed by the contour (i.e., the value $k$ for which the predicate $l_C^k$ is true is calculate for each line \verb+l[m]+). For example, the line $l =(2.5,0)= ((2,0),(3,0))$ is a bridge, being crossed by an equal number of left and right moves; this information is calculated adding positions 9 and 10 in the sorted list.\svsp

4. Conditions \ref{closedshape}, \ref{repeatparse} and \ref{intersections} are checked at lines 28-31, 22-23 and 24-27 respectively.
\eex

With minor adjustments, the procedure above may generate a normal form of the input contour. 

\subsection{The 2nd algorithm - an optimized version}

This version is an optimized version of the 1st algorithm. It applies a computational geometry technique (line sweeping) adapted to the set of segments composing a contour. Each letter repetition denotes a longer vertical or horizontal segment with a given orientation and length.  The time complexity of the algorithm reduces to $O(nr \log{(\max{y})})$, where $nr$ is the number of segments and $\max(y)$ is the difference between the greatest and the lowest $y$ coordinate reached.

{\it Sweep line or sweep surface} is a common concept in geometric algorithms. Usually, it consists in a vertical imaginary line that moves across the plane and stops in certain points, where its state is changed. The solution is found after all the stop points (events) are processed, gathering informations from all the neighboring objects. 

In the case of a contour, {\em stop points}, sorted in increasing order, are all $x$-coordinates of the composing segments. Hence there are three possible events: 
\begin{itemize}
\item left margin $(x1,y)$ of a horizontal segment ($l$ or $r$) $\Rightarrow \ Update(y,+1)$; 
\item right margin $(x2,y)$ of a horizontal segment $\Rightarrow \ Update(y,-1)$;  
\item vertical segment $(x,y1,y2)$ ($u$ or $d$) $\Rightarrow \ Query(y1,y2)$.
\end{itemize} 

A balanced binary tree may store the {\em line state}. Each node corresponds to an interval $[y1,y2]$ with offspring $[y1, (y1+y2)/2]$ and $[(y1+y2)/2 + 1, y2]$. The information $AUX[y1,y2]$, memorized as a heap, indicates the number of horizontal arrows intersected by the sweep line between $y1$ and $y2$, meaning how many $(x1,y), y\in [y1,y2]$ were swept without the corresponding $(x2,y)$ to by reached. Events of first and second type update the interval tree, adding or subtracting 1 to a certain leaf $AUX[k]$, where $k$ is the heap index corresponding to an interval of size $0, [y,y]$. $AUX[k][0], AUX[k][1]$ counts the number of segments oriented left and right, respectively, reaching the coordinate $y$. All nodes on the path from root (corresponding to the interval $[0, max(y)]$), to the leaf $k$ are updated.

The algorithm detects possible self-intersections when reaching a vertical segment $[y1,y2]$. By questioning the line state it verifies that no horizontal segments lies between $[y1+1, y2-1]$. $Query(y1,y2)$ is a divide and conquer procedure that sums informations found in the set of vertices composing a minimal partition of the segment $[y1,y2]$. If a  query returns at least 1 then the contour is self intersecting.

The length of the root interval of the balanced tree, determines the complexity of each update and query operation:  $O(\log{(\max(y)})$. As the number of events can't exceed twice the number of segments, the overall complexity is  $O(nr \log{(max(y)})$.      

\begin{figure}[h]
\centering
\begin{small}\begin{tt}\begin{algorithmic}[1]
  \Function{Query}{int index, int rootLeft, int rootRight, int qLeft, int qRight}
 \State \textbf{If }{rootLeft $\ge$ qLeft and rootRight $\le$ qRight} 
\State \hspace{0.15cm} {\textbf{return }abs(AUX[index][0] - AUX[index][1]);}  
 \State int resultLeft = 0, resultRight = 0, m = (rootLeft+rootRight)/2 \;
\State  \textbf{If }{leftQ $\le$ m} 
\State \hspace{0.15cm}  {resultLeft = Query(index*2,rootLeft,m,qLeft,qRight);}  
\State  \textbf{If }{rightQ $>$ m} 
\State \hspace{0.15cm}  {resultRight = Query(index*2+1,m+1,rootRight,qLeft,qRight) ;}  
\State \textbf{return} (resultLeft + resultRight) ;
  \EndFunction
\end{algorithmic}
\end{tt}\end{small}
\caption{An optimized version of the Algorithm in Fig.~\ref{alg:sort} - the key function}
\end{figure}

\bex The contour $C=r^4dlu^2rdl^2d^2lu^3ld$  
\begin{center}\includegraphics[scale=.15]{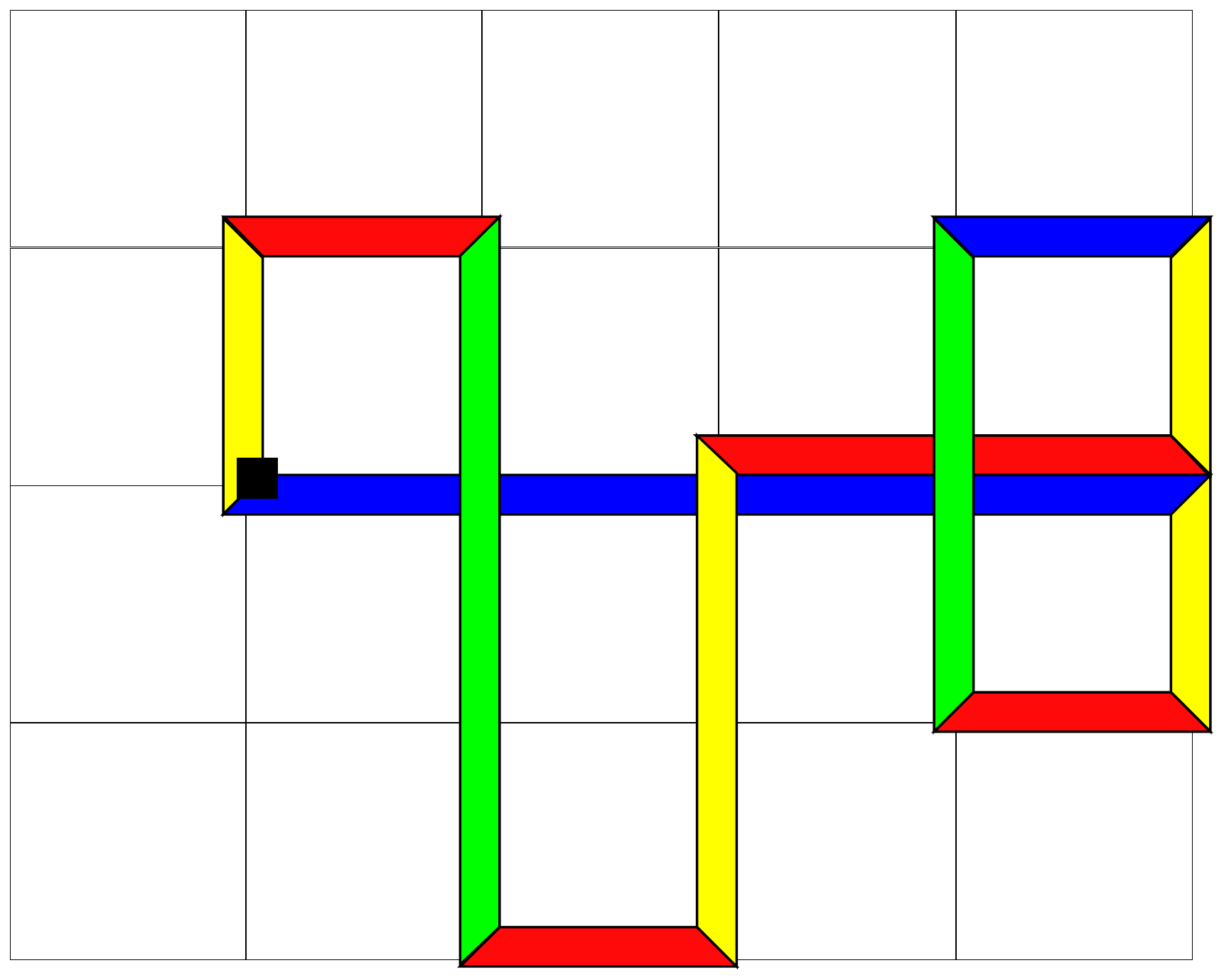}\end{center}
generates the list of events\svsp\\
\hspace*{.5cm}\parbox[c]{13cm}
{$U(2,+l)$, $U(3,+r)$, $U(0,+l)$, $Q(1,2)$ = 1, $U(3,-l)$ ,  $U(2,+l)$, $Q(1,1)$ = 0, $U(0,-l)$, $U(1,+l)$, $U(3,+r)$, $Q(2,2)$ = 0, $U(1,-l)$, $U(3,-l)$.}
\svsp\\
The first query event corresponds to the first vertical segment, considering that segments are ordered according to $x$ coordinate. At this point, the state of the sweep-line memorize one horizontal segment oriented to right, with $y$ in the query interval $[1,2]$. Hence $Q(1,2) = 1$.
\eex

\subsection{Comparison of the algorithms}
For the comparison of the the two algorithms presented above, we have performed tests on nine sets of randomly generated contours, of various length and shapes. 

We first analyse the performance on dense contours, consisting in one edge-connected component. The average execution time, in milliseconds, for each set of valid dense contours is presented in Table \ref{table:cdense}. The last column shows, in percent, the time of the 2nd algorithm compared to the 1st (the largest the contours, the better the 2st algorithm improvement).

The results obtained for sparse contours is summarized in Table \ref{table:csparse}. Sparse contours are valid composites of small rectangular contours with identities of various length. It may be seen that better execution times are obtained with the 2nd algorithm in case of contours with large distances between components or with components represented with large segments. 

Finally, results for composed contours mixing the above two sets are shown in Table \ref{table:ccomposed}. The notation DENSE$i$/SPARSE$i$ refers to the $i$-th line in the DENSE/SPARSE table above.

\begin{table}[tbh]
\begin{tabular}{| c | c | c | c |}
\hline
{\em DENSE} & \multicolumn{2}{|c|}{Average execution time} & time T2 / time T1 (\%)\\
\hline
contour length & first algorithm T1 & second algorithm T2 & T2*100/T1 \\
\hline
$<10000$ & $0,3572$ &  $0,1049$ & $29,8947$\\
$>10000$ & $2,0573$ & $0,4015$ & $20,1360$\\
$<500$ & $0,0688$ & $0,0667$ & $94,3219$\\
\hline
\end{tabular}
\caption{Comparison of the algorithms on dense contours}
\label{table:cdense}
\end{table}

\begin{table}
\begin{tabular}{| c | c | c | c | c |}
\hline
\multicolumn{2}{|c|}{\em SPARSE} & \multicolumn{2}{|c|}{Average execution time} & time T2 / time T1 (\%)\\
\hline 
cells & distance  & first algorithm T1 & second algorithm T2 & T2*100/T1 \\
\hline
$[100,200]$ & $[25,50]$ &  $2,6484$ &  $1,2521$ & $47,3936$\\
$[200,300]$ &$[25,50]$ & $3,7114$ & $1,7215$ & $46,4073$\\
$[100,200]$ &$[50,100]$ & $7,1069$ & $2,5222$ & $35,5811$\\
\hline
\end{tabular}
\caption{Comparison of the algorithms on sparse contours}
\label{table:csparse}
\end{table}

\begin{table}
\begin{tabular}{| c | c | c | c | c |}
\hline
\multicolumn{2}{|c|}{\em COMPOSED} & \multicolumn{2}{|c|}{Average execution time on c1.c2} & time T2 / time T1 (\%)\\
\hline 
c1 & c2  & first algorithm T1 & second algorithm T2 & T2*100/T1 \\
\hline
DENSE1 & SPARSE1 &  $2,9360$ &  $0,9170$ & $31,3385$\\
DENSE2 & SPARSE2 & $6,9120$ & $1,7882$ & $26,2817$\\
DENSE3 & SPARSE3 & $5,8749$ & $1,6525$ & $28,2066$\\
\hline
\end{tabular}
\caption{Comparison of the algorithms on composed contours}
\label{table:ccomposed}
\end{table}

We mention that in both implementations, we stop when one of the conditions \ref{closedshape}, \ref{repeatparse} or \ref{intersections} is not checked. Hence, for invalid contours, the performance my also depend on the start point, the place where the first self-intersection is placed, etc. 

\section{Conclusions and future works}

The known approach \cite{ga-ra92} to get regular expressions for 2-dimensional patterns uses intersection and renaming - see \cite{koz91,bps13} for some critics on using these operators. One of the benefits of our approach here and of the new type of regular expressions n2RE introduced in \cite{bps13} is that renaming and intersection are avoided, the setting being closer in spirit with classical 1-dimensional regular expressions.

Current hardware and software development, mainly driven by multi-core architectures and distributed computing technologies, bring forward the necessity to adapt sequential machine models to interactive computation. The results in this paper are steps of a program extending sequential computation models in this direction. 

As future work we intend to prove a Kleene theorem for finite interactive systems and also to develop an associated algebraic theory, similar to automata theory. Possible applications of the model are: image processing and image recognition procedures, study of parallel, interactive OO-programs, modelling discrete physical or biological systems etc.


\begin{thebibliography}{99}

\bibitem{acm02}E. Asarin, P. Caspi, and O. Maler. Timed regular expressions. {\em Journal of the ACM}, 49:172--206, 2002.
\bibitem{bps13}I.T. Banu-Demergian, C.I. Paduraru, and G. Stefanescu.  A new representation of two-dimensional patterns and applications to interactive programming. In: {\it Proceedings FSEN 2013}, LNCS 8161, pp. 183--198.  Springer, 2013. 
\bibitem{ben79} J. Bentley and T. A. Ottmann. Algorithms for reporting and counting geometric intersections. {\it Computers, IEEE Transactions}, 100(9):643--647, 1979.
\bibitem{bib01} E. Bribiesca and C. Verlade. A formal language approach for a 3D curve representation. {\it Computers \& Mathematics with Applications} 42(12):1571--1584, 2001.
\bibitem{conway71}J.H. Conway. {\em Regular Algebra and Finite Machines}. Chapman and Hall, 1971.
\bibitem{dr-st08b}C.~Dragoi and G.~Stefanescu. AGAPIA v0.1: A programming language for interactive systems and its typing systems. In: {\it Proc. FINCO/ETAPS 2007}, ENTCS 203, pp. 69-94. Elsevier, 2008.
\bibitem{free} H. Freeman. On the encoding of arbitrary geometric configurations. {\it Electronic Computers, IRE Transactions} 2:260--268, 1961.
\bibitem{ga-ra92}V.~Garg and M.T. Ragunath. Concurrent regular expressions and their relationship to {P}etri nets. {\em Theoretical Computer Science}, 96:285--304, 1992.
\bibitem{gi-re97}D. Giammarresi and A. Restivo A. Two-dimensional languages. In: {\em Handbook of Formal Languages. Vol.~3: Beyond Words}, 215--265. Springer-Verlag, 1997.
\bibitem{gsv06}D.~Goldin, S.~Smolka, P.~Wegner (Eds.). {\em Interactive Computation: The New Paradigm.} Springer, 2006.
\bibitem{kle56}S.C. Kleene. Representation of events in nerve nets and finite automata. In: {\em Automata Studies}, 3--41. Princeton University Press, 1956.
\bibitem{koz91}D.~Kozen. A completeness theorem for Kleene algebras and the algebra of regular events. In: {\em Proc.~LICS 1991}, 214-225.
\bibitem{koku85}T. Kaneko and M. Okudaira. Encoding of arbitrary curves based on the chain code representation. {\it Communications, IEEE Transactions}, 33(7):697--707, 1985
\bibitem{liu05}Y. K. Liu and B. \u{Z}alik. An efficient chain code with Huffman coding. {\it Pattern Recognition}, 38:553--557, 2005.
\bibitem{ste01a}G.~Stefanescu. Algebra of networks: Modeling simple networks as well as complex interactive systems. In: {\it Proof and System Reliability} 49--78. Kluwer, 2002.
\bibitem{ste06a}G.~Stefanescu. Interactive systems with registers and voices. {\it Fundamenta Informaticae}, 73:285-306, 2006.
\end{thebibliography}
\end{document}